# Boron films produced by high energy Pulsed Laser Deposition

D. Dellasega [a,*], V. Russo [a], A. Pezzoli [a], C. Conti [b], N. Lecis [c], E. Besozzi [a], M. Beghi [a], C.E. Bottani [a], M. Passoni [a]

[a] Department of Energy, Politecnico di Milano, via Ponzio 34/3, 20133 Milan, Italy
[b] ICVBC, National Research Council, Via Cozzi 53, 20125 Milan, Italy
[c] Department of Mechanical Engineering, Politecnico di Milano, via La Masa 1, 20156 Milan, Italy

Abstract

Micron-thick boron films have been deposited by Pulsed Laser Deposition in vacuum on several substrates at room temperature. The use of high energy pulses (>700 mJ) results in the deposition of smooth coatings with low oxygen uptake even at base pressures of $10^{-4}$-$10^{-3}$ Pa. A detailed structural analysis, by X-Ray Diffraction and Raman, allowed to assess the amorphous nature of the deposited films as well as to determine the base pressure that prevents boron oxide formation. In addition the crystallization dynamics has been characterized showing that film crystallinity already improves at relatively low temperatures (800 °C). Elastic properties of the boron films have been determined by Brillouin spectroscopy. Finally, micro-hardness tests have been used to explore cohesion and hardness of B films deposited on aluminum, silicon and alumina. The reported deposition strategy allows the growth of reliable boron coatings paving the way for their use in many technology fields.

## 1. Introduction

Boron (B) and boron compounds (e.g. $B_xO$, $B_4C$, B-Al, $MgB_2$) exhibit unique and very interesting properties that find application in various technology fields. Thanks to their extreme hardness, up to 30–60 GPa [1], B films can be exploited as protective coatings [2]. When exposed to ambient air, B surfaces get oxidized forming a hydroxide layer (boric acid), which acts as a solid lubricant that allows high wear properties as well as antibacterial properties [2, 3]. Boron surfaces exhibit high reflectance in the extreme ultraviolet (40 nm - 200 nm), and are therefore suitable for the production of extreme UV optics that find applications in plasma diagnostics, spectroscopy, synchrotron radiation, and free electron lasers [4]. Boron-based materials are also used in the framework of thermoelectric energy conversion studies [5]. Coatings containing the isotope boron-10 ($^{10}$B), one of the two stable isotopes of boron, are widely studied in the field of neutron detection thanks to the high cross–section of the $^{10}$B(n, α)$^7$Li nuclear reaction. Moreover boron is often incorporated in neutron shields, taking advantage of its high cross-section for neutron capture [6]. Thanks to its affinity with oxygen, boron has been used in magnetic nuclear fusion research as wall conditioning on many tokamaks, achieving in this way a remarkable improvement on the plasma performances [7]. Among the boron compounds,

MgB$_2$ is a very extensively studied superconductor material [8]. All these features are related to its very peculiar crystalline structure. Boron atoms are grouped in icosahedra (B$_{12}$), which in turn are orderly arranged in a lattice. Crystalline B presents four main crystallographic phases: α-rhombohedral where B$_{12}$ icosahedra are positioned at each of the eight vertices of the rhombohedral unit cell, β-rhombohedral where B$_{12}$ are present also on the edges of the cell. Gamma orthorhombic phase has been synthesized at high pressures. It shows a rock-salt type arrangement of the icosahedra. A tetragonal phase is also present but foreign atoms are required to ensure its stability [9]. Also in its amorphous phase, in which the long range order is lost, the short range order of the B$_{12}$ icosahedra is preserved [9, 10]. Very recently exotic structures where B is not arranged in B$_{12}$ units such as boron nanowires [11], boron fullerene (B$_{40}$ molecules) [12] and borophene [13] (graphene-like structures composed of boron atoms) have been synthesized.

Application of pure boron in many technology fields is presently limited due to the extreme difficulty in depositing high quality B films or coatings. Boron cannot be electroplated [2]. It is possible to deposit B film by Chemical Vapor Deposition, but the needed feed gases are expensive and often toxic and/or explosive [14]. Among the Physical Vapor Deposition techniques, boron carbide films are nicely produced by magnetron sputtering [15, 16], but pure B is difficult to sputter because of its low atomic number [17]. Boron films have been deposited by e-beam evaporation in form of very thin films (~ 100 nm) with very low deposition rates (0.06–0.58 nm/s) [18]. Cathodic arc deposition allows the production of compact B coatings thanks to the high energy of the depositing species. A careful tailoring of the process parameters is needed because, due to their brittle nature, B targets do not withstand the thermal loads applied during the operation and can disintegrate [2].

Pulsed Laser Deposition (PLD), thanks to its out of equilibrium deposition mechanisms, has been proven to be able to deposit several kinds of coatings that are difficult or impossible to deposit with other techniques, like e.g. YBCO [19], Ag$_4$O$_4$ [20], amorphous-like W [21, 22] and C foam with ultra–low density [23]. This kind of technique has also been exploited for the deposition of B-based coatings for several applications such as MgB$_2$ superconducting films [8, 24, 25], Boron sub-oxide (B$_6$O) [26, 27] and AlMgB$_{14}$ [28] as ultra–hard coatings. There are also few works focused on the deposition of pure Boron films in the framework of MgB$_2$ production [8] and to be used as thermal-neutron converters [29]. These studies are aimed at characterizing, mainly by XPS, surface stoichiometry of B films. It is interesting to mention that, thanks to the features of the PLD process, $^{10}$B enriched boron carbide films have been obtained. $^{10}$B/$^{11}$B ratio of 0.9, against the natural abundance of 0.25, is found at a laser fluence of 6.4 J/cm$^2$ [30]. Enriched B$^{10}$ films can be used for the neutron detection. One of the main issues in the deposition of boron coatings, is the presence of an uncontrolled amount of oxygen in the growing film that usually degrades film properties. In the work of Mijatovic et al. [8] oxygen content in the growing film has been found to be critically dependent on the background pressure and purity of the ambient gas. It is worth noting that the PLD process performed in a tubular furnace at high temperature (700 °C – 1200 °C) leads to the deposition of very pure boron films (Oxygen content of 5% or lower) [31]. But except for this very peculiar setup the oxygen uptake is one of the major open issues in the B coatings deposition. In addition a full mechanical and structural characterization of the boron films deposited by PLD cannot be easily found.

Usually in PLD experiments a low value of laser fluence (~1 J/cm$^2$) is adopted with the purpose of limiting the amount of ejected droplets [32]. On the other hand recently we have shown that a high pulse energy (> 550 mJ) and high fluence (> 8 J/cm$^2$) allow the production of coatings with new and interesting properties. For example high pulse energy and fluence (850 mJ and 8.8 J/cm$^2$ respectively) have been found to induce the deposition of amorphous-like W films inhibiting the crystalline growth during the synthesis [21]. In addition, droplet-free Rh films have been deposited by PLD using laser pulses of 1700 mJ at a laser fluence of 19 J/cm$^2$ thanks to the high energy of the plasma plume [33].

In the present work we show how the use of high pulse energy and high fluence (E > 700 mJ and 9 J/cm$^2$ respectively) in the nanosecond regime allows to deposit dense, micron thick B coatings on different substrates (Si, Al and alumina) at room temperature as well as to control and significantly decrease the oxygen content in the growing boron film. Laser pulse energy and fluence have been related to the film morphology and oxygen content in the deposited coating. The most significant features of the deposited films have been extensively characterized: morphology (by Scanning Electron Microscopy), elemental composition (by Energy Dispersion Spectroscopy), crystalline structure (by X-Ray Diffraction and Raman Spectroscopy) and mechanical properties (by Brillouin Spectroscopy and Micro-Hardness Tests). In addition the capability of PLD of producing micrometer thick coatings on reliable areas with controlled morphology ready for "realistic" applications is shown. This will foresee the use of boron coatings in many technological fields, as an example in the bio-medical area where hard, biocompatible and, in the present case even antibacterial coatings are of extreme interest.

## 2. Experimental details

In our experimental setup 532 nm laser pulses at 7 ns are partially focused on a 2 inch diameter boron target. The pulse energy varies between 200 mJ and 800 mJ, resulting in laser fluence ranging from 2.3

J/cm$^2$ to 9 J/cm$^2$, the laser spot area being of 8.8 mm$^2$. The atoms ablated from the target expand in a vacuum chamber whose base pressure is varied between 5*10$^{-4}$ Pa and 8*10$^{-3}$ Pa. The expanding species are collected on silicon (100), aluminum or alumina substrates positioned 6 cm or 9 cm away from the target. Silicon substrates are single side polished with an average roughness of 2 nm. Aluminum is laminated with a surface roughness of about 400 nm. Alumina substrates are made of sintered alumina powder with size 3-10 μm. In this case the average roughness is 1.5 μm – 2 μm. The B coatings deposited on alumina have been subsequently annealed in vacuum (10$^{-5}$ Pa) at 800°C and 1200°C for 1 h. Film structure have been characterized by X-Ray Diffraction (XRD), using a Panalytical X'Pert PRO X-ray diffractometer in θ/2θ configuration. K$_α$ radiation have been used, current at the anode is 40 mA with a tension of 40 kV. The investigated range is 9.99 -64.99 degrees, the step size is 0.0167113 degrees and time at each point is 10.8 s. Film morphology have been investigated by a Zeiss Supra 40 Field Emission Scanning Electron Microscope (SEM, accelerating voltage 3-5 kV). In order to check film composition, we performed Energy Dispersion Spectroscopy analysis (EDS) using an accelerating voltage of 5 or 6 kV, to excite K$_α$ electronic levels of B, C, N, O, Mg and Si. EDS system has been calibrated using Si. Nevertheless due to the low Z of the elements involved this kind of analysis can be considered only semi-quantitative in indicating the absolute composition. Stoichiometry and crystallinity of the samples have been investigated by Raman Spectroscopy using a Renishaw InVia spectrometer in backscattering geometry. The excitation wavelength is 514.5 nm from an Ar ion laser. Film density has been estimated using a quartz crystal microbalance and SEM cross section. Mechanical properties of boron films are measured by Surface Brillouin Spectroscopy (SBS). A Nd:YAG continuum laser operating at 200 mW at a wavelength λ = 532 nm is focused on the free surface of the film. The spot size on the sample is of tens of micrometers. The backscattering geometry is adopted: the exchanged wave-vector of the surface acoustic waves, parallel to the surface, is thus determined by the incidence angle θ, which is varied from 60° to 40°, as $k_\parallel = 2(2\pi/\lambda_0)sin\theta$. The sample is oriented to investigate the propagation of the Surface Acoustic Waves along the Si[110] direction. The scattered light is collected and analyzed by a tandem multi–pass Fabry-Perot interferometer of the Sandercock type with a free spectral range typically of 50 GHz. More details on the full setup can be found in [34].

The micro–hardness tests were performed using a Instrumented micro-indenter (Microcombi Platform CSM Instrument), equipped with a Vickers square based pyramidal diamond indenter, load controlled, applying variable loads of 50, 100, 200, 300 mN. Correspondingly loading and unloading rates have been set at: 100, 200, 400 and 600 mN/min. No pretreatment have been made on the coatings before the tests.

## 3. Results and Discussion

### 3.1. Morphology and oxygen content

In Fig.1 SEM images of B films deposited on silicon at $5*10^{-4}$ Pa at various pulse energies/fluences are shown. It is evident that these parameters deeply influence the nanostructure of the growing film. At 200 mJ (corresponding fluence 2.3 J/cm$^2$), Fig.1a, the film surface is very rough and composed by micro– and nano-aggregates. At about 450 mJ (corresponding fluence 5.1 J/cm$^2$), Fig.1b, film surface becomes smooth, with a great reduction of surface defects (indicated by white arrows). At 800 mJ (corresponding fluence 9.1 J/cm$^2$), Fig.1c the defect number is even smaller, showing a smooth surface. Cross section analysis related to the film deposited at the highest pulse energy, Fig.1d, shows a featureless morphology where some cauliflower growths are also visible. At lower magnifications, (as visible in Figs 8a and 8b) we found that the film surface is characterized by the presence of sub-micrometer size aggregates, probably due to the deposition of molten droplets coming from the target during the ablation process. Raising the pulse energy results also in a reduction of the micrometer defects.

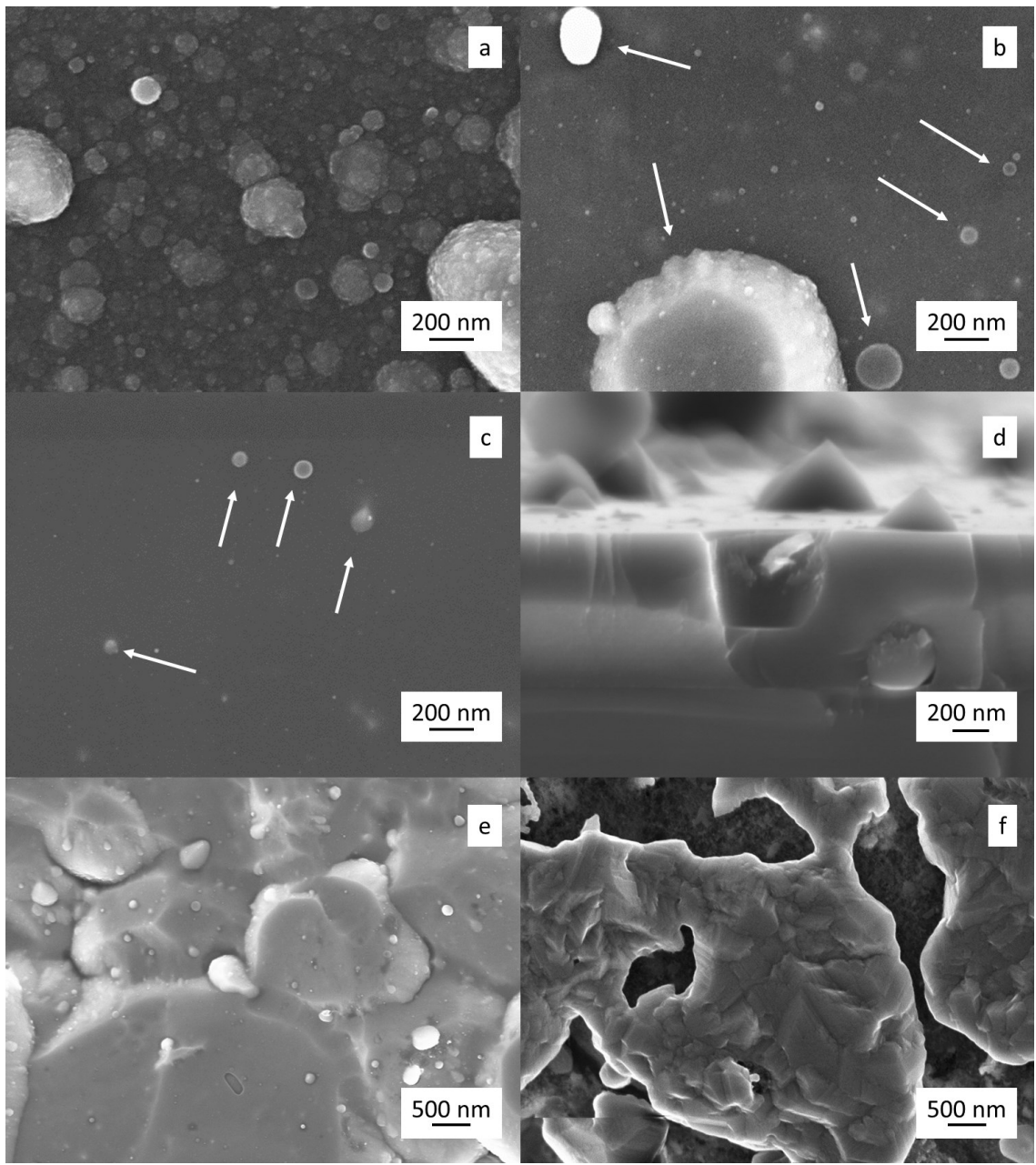

Figure 1: SEM images of boron films deposited on silicon at 5*10$^{-4}$ Pa with pulse energy of: a) 200 mJ , b) 450 mJ , c) 800 mJ, plain view and d) 800 mJ, cross section. B films deposited on sintered alumina at 800 mJ and annealed in vacuum at: e) 800 °C and f) 1200 °C. Defects are indicated by white arrows.

|  |  | **B** | **C** | **N** | **O** | **Mg** | **Si** |
|---|---|---|---|---|---|---|---|
| B target | Mean | 84.9 % | 4.6 % | 0.2 % | 8.5 % | 1.6 % | 0.3 % |
|  | St. Dev. | 1.4 | 0.6 | 0.3 | 1.2 | 0.4 | 0.0 |
|  |  |  |  |  |  |  |  |
| B film ($10^{-4}$ Pa, 800 mJ) | Mean | 86.5 % | 0.0 % | 1.2 % | 10.5 % | 1.6 % | 0.2 % |
|  | St. Dev. | 0.1 | 0.0 | 0.1 | 0.1 | 0.0 | 0.0 |
|  |  |  |  |  |  |  |  |
|  | Δ (film-target) | **1.6 %** | **-4.6 %** | **1.0 %** | **2.0 %** | **0.0 %** | **-0.1 %** |

Table 1: EDS analysis of B target and B film deposited at $5*10^{-4}$ Pa and 800 mJ. Variation of the relative abundance of the single elements is also displayed.

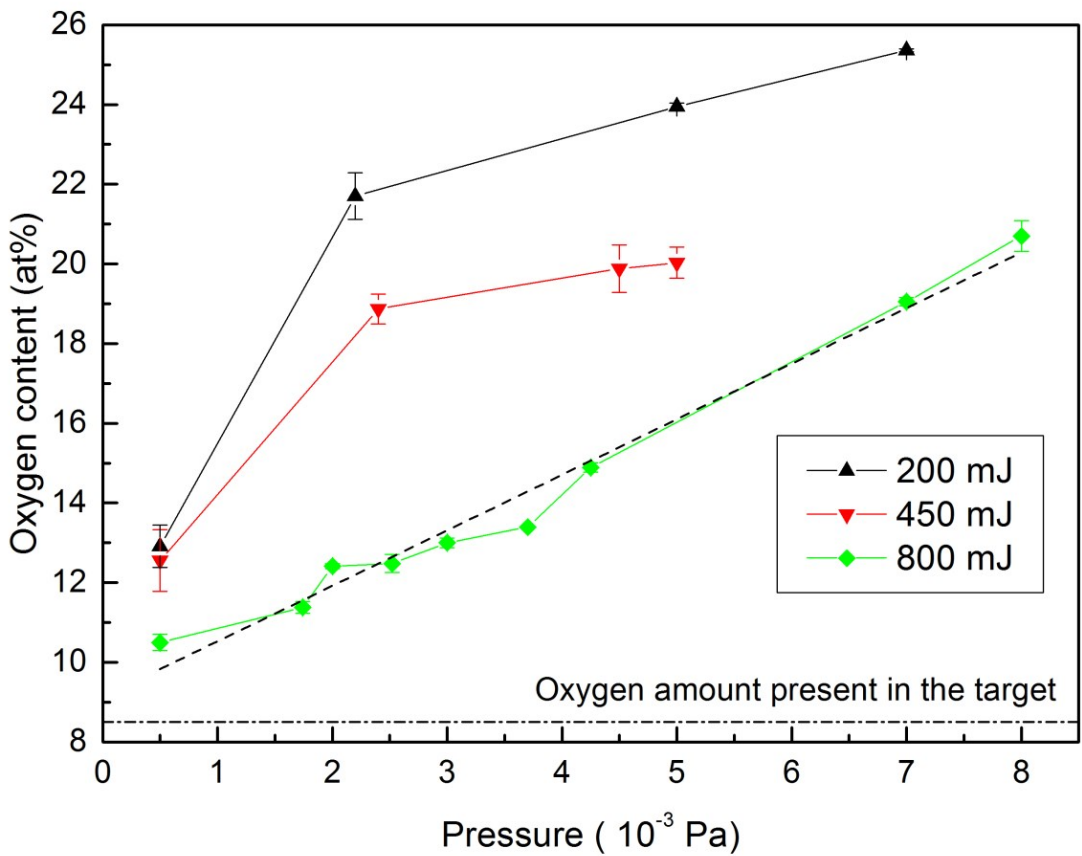

Figure 2: Oxygen content, measured by EDS, of the deposited B films versus deposition pressure varying pulse energy. The dotted line is the linear fit of the 800 mJ trend.

Pulse energy do not only influences film morphology but also affects the amount of oxygen retained in the growing film. In Table 1 the EDS analysis of both B target and B film deposited at $5*10^{-4}$ Pa and 800 mJ is presented. In the B target some impurities are present: C, N, O coming from environmental contamination, Mg and Si probably coming from the preparation procedures of the target. The most abundant impurities are O (8.5 %) and C (4.6 %). The composition of the deposited B coating is almost equal to the one of the B target. One of the main advantages of PLD is, in fact, the capability of replicating target stoichiometry during the film growth. Nevertheless some differences can be noted. B film is free of C, this superficial contamination is probably eliminated after the first laser shots. There is a slight incorporation of O and N in the deposited film whose amount increase by 2% and 1% respectively. These two elements are present in the residual atmosphere of the deposition chamber. Mg and Si impurities are transferred from target to substrate without significant variations. The incorporation of oxygen in the B films is critically dependent on the deposition conditions (i.e. base pressure and pulse energy).

In Fig.2 the oxygen concentration versus deposition pressure for different pulse energies is shown. Raising background pressure the oxygen content in the B films increases. At the lowest investigated pressure, $5*10^{-4}$ Pa, the oxygen content is 12.9%, 12.6% and 10.5% for the films deposited at 200 mJ, 450 mJ and 800 mJ respectively. At higher background pressures the oxygen content is even more dependent on the pulse energy. For the films deposited at 200 mJ and 450 mJ per pulse the oxygen content rapidly increases with base pressure reaching about 20% at $2*10^{-3}$ Pa. Above this pressure the slope of the curve changes and the oxygen amount grows in a more gradual way. At 800 mJ per pulse the oxygen content versus pressure grows almost linearly, being 12.5% at $2*10^{-3}$ Pa and remaining always much lower than the corresponding values found at lower energies. In the other few PLD experiments reported in literature O amount has been only related to the base pressure of the deposition system. Deposited B films show an O concentration of 11% [8] and 48% [29], the corresponding base pressures being of $5*10^{-7}$ Pa and $4*10^{-5}$ Pa respectively. On the contrary, in our experiment we show that pulse energy plays a fundamental role in the deposition of B films, determining a low oxygen content even at a relatively high base pressure ($5*10^{-4}$ Pa). The lower oxygen uptake during deposition at high energy pulses may be mainly due to the higher kinetic energy of the ablated species and the higher instantaneous deposition rate [21].

An additional oxygen depletion has been achieved by vacuum annealing. B films deposited on alumina with the lowest oxygen content (10.5%) have been annealed in vacuum at 1200 °C for 1 h. After the annealing treatment at the highest temperature the oxygen content in the film (characterized by EDS), is found to be decreased from the initial value of 10.5% to 0.5%. Fig. 1f presents the SEM image of B coatings after the thermal treatments. At 1200 °C the formation of crystalline features can be appreciated. Since the crystallization temperature of amorphous boron is about 1160 °C [35], the phase transition from amorphous to crystalline plays a pivotal role in removing the remaining oxygen from boron film.

We found that both morphology of the deposited boron coatings as well as the boron purity improve raising pulse energy. Thus we concentrated our characterizations on the coatings produced at the highest pulse energy/fluence (800 mJ, 9 J/cm$^2$)

**3.2 Crystalline structure and stoichiometry**

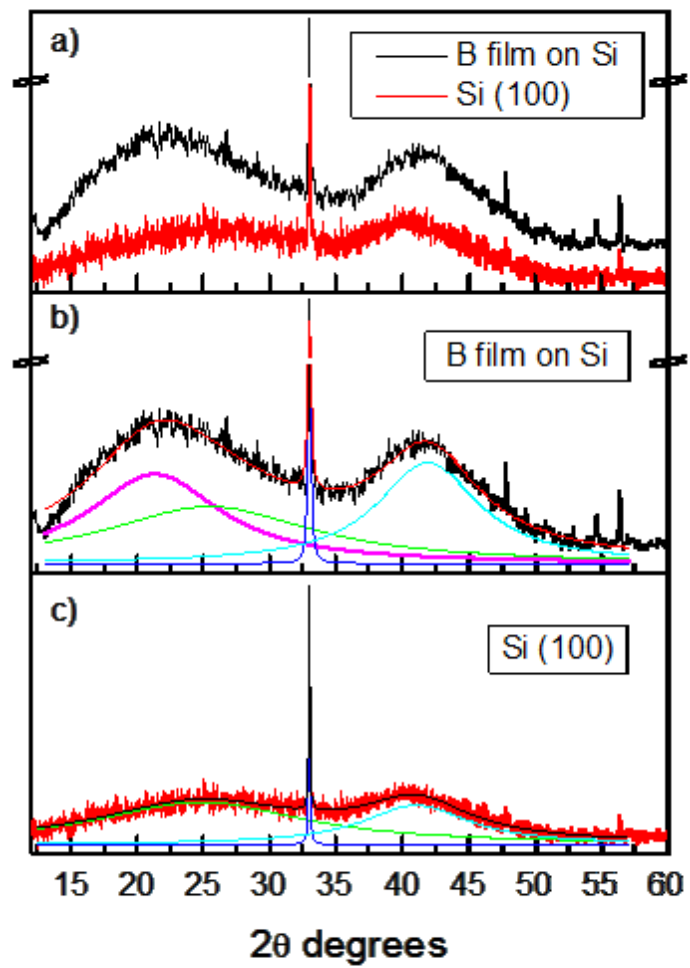

Figure 3: XRD analysis of B film and Si substrate. a) reflections of boron film on Si, and of bare Si substrate. b) lorentzian peak fitting of the XRD spectrum of B on Si. c) peak fitting of bare Si substrate.

Cyan, blue and green fitting curves are related to silicon substrate. Magenta fitting curve is related to boron coating.

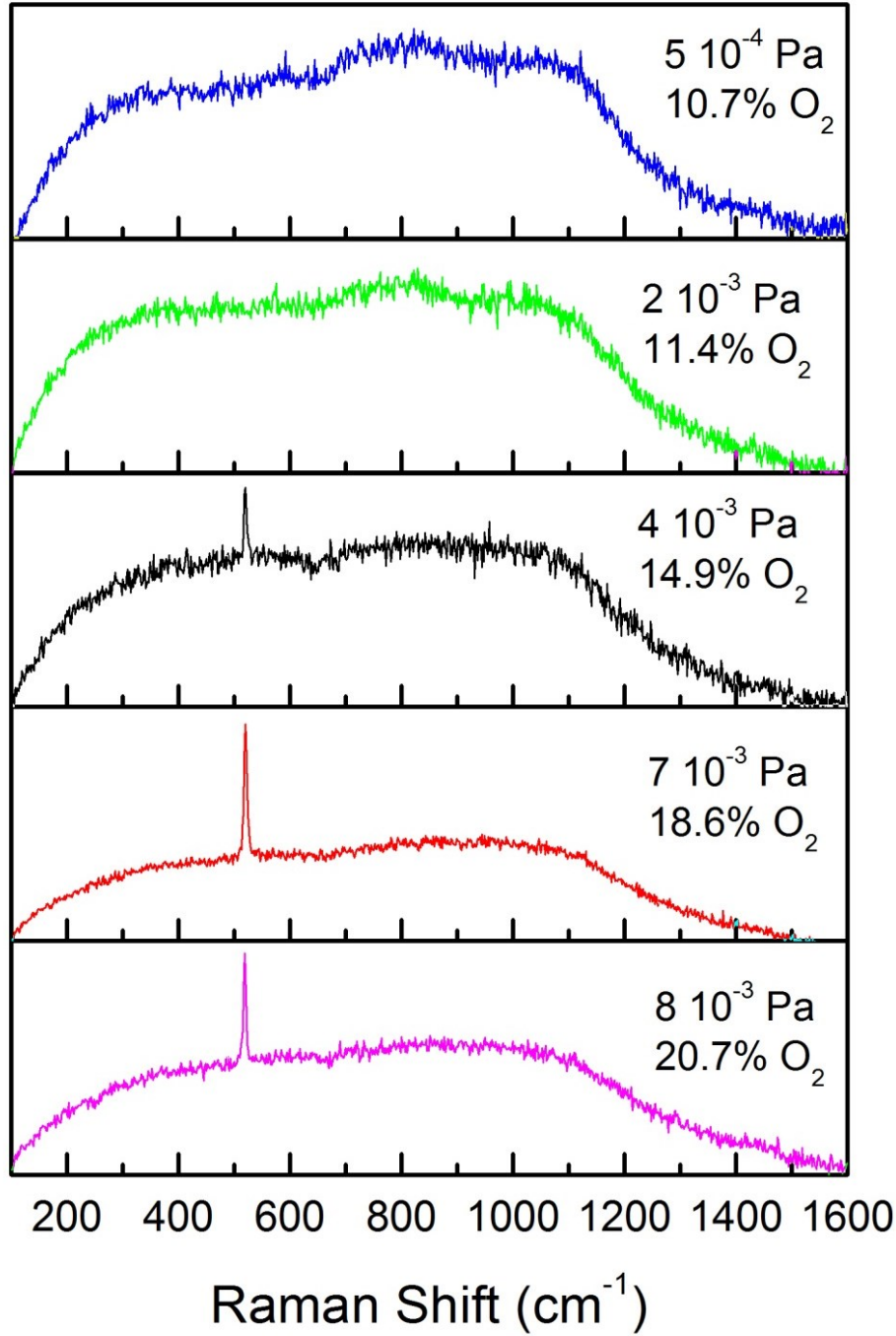

Figure 4: Raman analysis of B films deposited at 800 mJ varying deposition pressure from $5*10^{-4}$ Pa to $8*10^{-3}$ Pa.

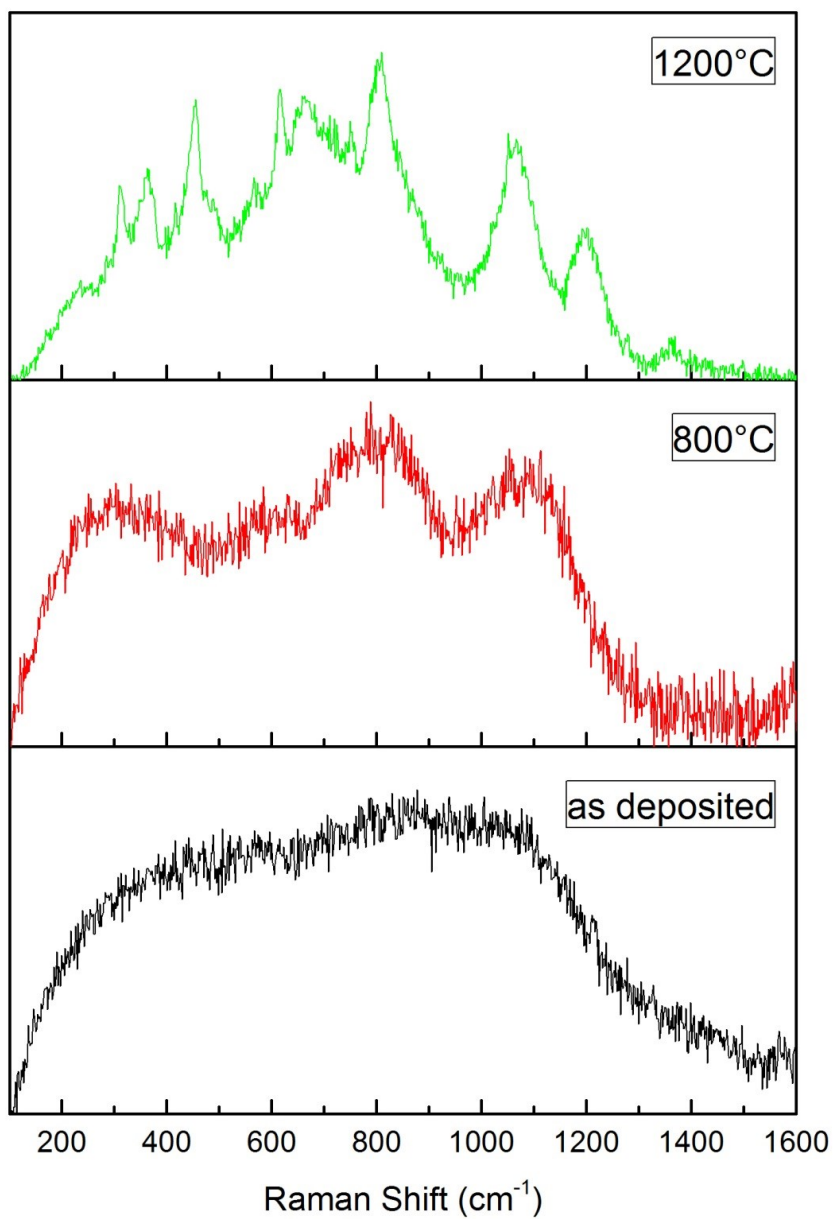

Figure 5: Raman analysis of B films deposited at 800 mJ and 5*10$^{-4}$ Pa and annealed in vacuum at 800 °C and 1200 °C.

Microstructure and stoichiometry of the B films deposited at 800 mJ have been characterized by XRD and Raman spectroscopy. In Fig.3a the XRD spectrum of B film deposited on Si at 5*10$^{-4}$ Pa is shown. The recorded spectrum is constituted by a peak at 33° and two broad bands at about 21° and 42°. Reflections coming from the Si substrate greatly perturb the signal coming from the coating (Si spectrum is added for comparison). After a proper fitting of the reflections of bare Si samples, see Fig3 c, it is possible to argue that the contribution of Si substrate to the whole spectrum is related to two bands at 25.46° and 41.9° and a peak at 33.01°. As a result in the fitting of B deposited on Si, Fig 3b, the band at 21.3° may be related to the B film. The presence of a reflection in the range 20.17° – 21.14° is a typical feature of amorphous boron or amorphous B compounds (e.g. amorphous $B_6O$) that, despite their amorphous structure, maintain some degree of order related to the presence of icosahedral units [36, 37]. To better characterize structure and stoichiometry of the deposited films, Raman analysis of the B films deposited on Si at 800 mJ with different base pressures have been performed, see Fig.4. For all the investigated pressures, Raman spectrum is characterized by a wide band between 200 and 1200 $cm^{-1}$, confirming the amorphous nature of the films. This large band resembles the phonon density of states computed by Shirai et al. using a Bethe lattice of connected icosahedra [38] and it can be de-convoluted into different contributions by a fitting procedure (not shown). For any boron compounds it is possible to identify three principal spectral regions in the Raman spectrum related to vibrational modes: a region at high wavenumbers, over 1100 $cm^{-1}$, related to inter-icosahedral B-B vibrations, a region between 400 and 1100 $cm^{-1}$ related to intra-icosahedral B-B vibrations and a low wavenumber region, 200–400 $cm^{-1}$, attributed to modes involving chains possibly formed by icosahedra and intermediated atoms (B atoms, oxygen atoms or other atoms in boron compounds) [9, 38-39]. For lower pressures, up to $2*10^{-3}$ Pa (corresponding to an O content of 11.4%), only this large band is visible in the spectrum. Moving to higher pressure, the sharp peak at 521 $cm^{-1}$ due to silicon substrate appears, more and more intense with increasing pressure, revealing an increasing transparency of the film. This behavior can be attributed to incorporation of oxygen in the film with consequent formation of B–O covalent/ionic bonds, substituting covalent B-B bonds [40]. From this observation it is possible to define, within the used high pulse energy regime, a threshold pressure at about $2*10^{-3}$ Pa, below which amorphous pure B coating is deposited without reacting with oxygen. Starting from one of this pure boron coatings, deposited on alumina at $5*10^{-4}$ Pa, the effect of annealing treatments on the film structure and morphology is studied by Raman spectroscopy and SEM (see Figs 1e and 1f). In Fig.5 are shown Raman spectra of coatings as deposited and after annealing in vacuum at 800 °C and 1200 °C. The spectrum of as deposited B film is identical to the corresponding film deposited on silicon shown in fig.4. For the film annealed at 800 °C the spectrum evolves and four more defined bands can be identified at about 277 $cm^{-1}$, 600 $cm^{-1}$, 817 $cm^{-1}$ and 1075 $cm^{-1}$, suggesting a tendency towards ordering. The corresponding morphology does not show any modifications respect to the non-annealed sample, see Fig 1e (The surface roughness at the micrometer scale visible in fig 1e is related to the alumina substrate). The obtained spectrum reproduces quite well both Raman spectrum and calculated phonon DOS reported in ref. [38] that investigates the vibrational properties of icosahedron-based random network. It is possible to argue that this annealing temperature is sufficient to induce ordering of the single $B_{12}$ icosahedra but not enough to establish long-range order. When the film is annealed at 1200 °C crystallization is expected to occur [35]. SEM analysis shows the formation of crystalline facets on the film surface, see Fig. 1f. As a result the four Raman bands evolve and many peaks appear. The features of the spectrum are compatible with the nucleation of a β-rhombohedral crystalline phase with a very low content of impurities [9]. This crystallographic phase has been demonstrated to be the most stable at ambient conditions [41,42].

**3.3. Deposition on large areas, elastic properties and mechanical characterizations**

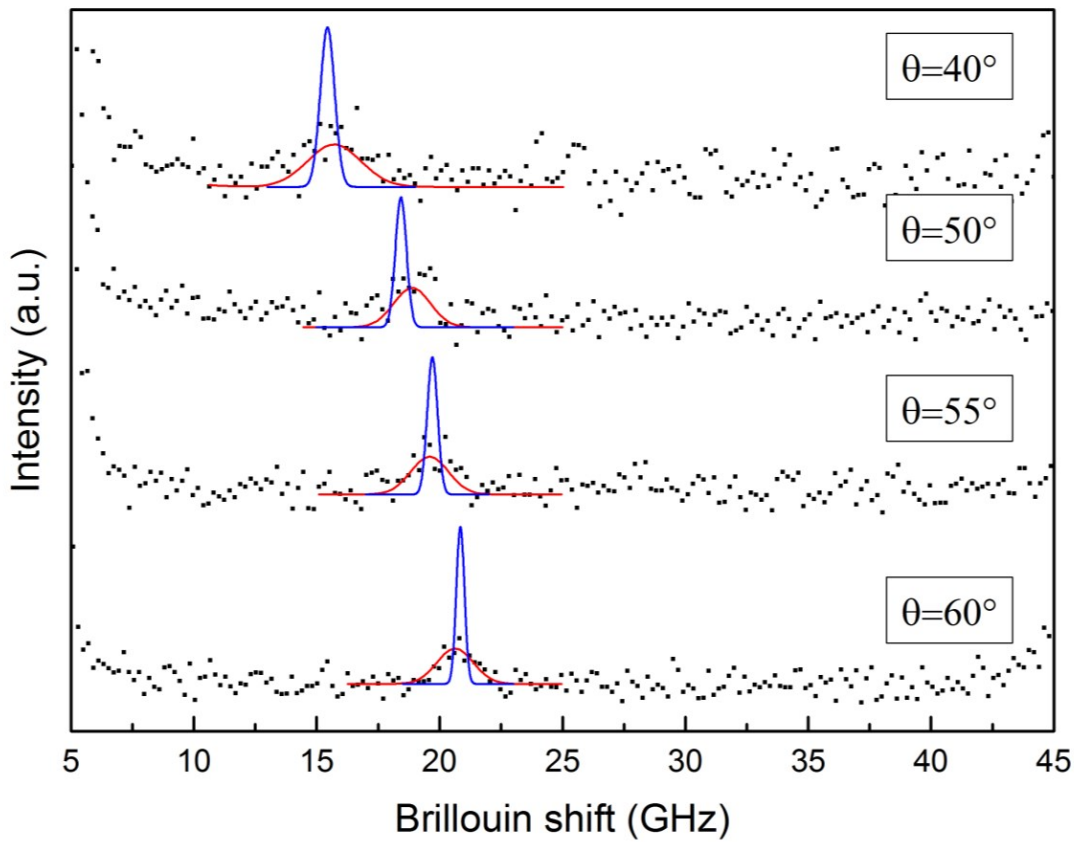

Figure 6: Brillouin characterization of B films deposited on silicon at 800 mJ and $5*10^{-4}$ Pa varying incidence angle. The Gaussian fit of the experimental data is reported in red. The theoretical peak frequencies obtained by the best fit parameters are reported in blue as Gaussian peaks. The width is estimated on the basis of scattering angle and the aperture of the scattered light collection. The area of every theoretical peak is set to the value of the experimental one.

In view of technological exploitations, it is important to show the ability to produce B coatings of reliable thickness on a sufficiently wide area and check the mechanical behavior of the coating on different substrates. Thanks to the use of a sample holder which keeps the substrate in motion during the deposition, and an off–axis deposition scheme [33], we have been able to deposit 0.6 μm - 1 μm thick boron coatings with lateral dimensions up to 40 mm, produced with pulse energy of 800 mJ, on different substrates, such as laminated aluminum, (100) crystallographically oriented polished silicon and sintered alumina substrates. The deposition rate ranges between 26 and 30 nm/min depending on substrate dimension. Film uniformity is about 6.5% on the whole surface. Film density has been assessed by the combined use of a quartz crystal microbalance and SEM cross section analysis as already reported in [43]. Mass density of the boron films grown at 800 mJ is 2.51 ± 0.35 g/cm$^3$. Although the obtained results are quite scattered the mean value is close to the mass density of bulk β-rhombohedral boron (2.35 g/cm$^3$).

The elastic properties of 600 nm thick amorphous boron films, deposited on silicon substrate at a pressure of 5*10$^{-4}$ Pa have been assessed by Surface Brillouin Spectroscopy (SBS). As far as the acoustic wave propagation is concerned, boron-silicon samples are fast films on slow substrates (i.e. stiffening layers), so that, at whatever thickness, the velocity of the Surface Acoustic Wave (SAW) is higher than the Rayleigh Wave (RW) velocity of the substrate. In this case the SAW in the presence of the layer is more appropriately classified as a pseudo-SAW (PSAW), because, being faster than the RW of the substrate it radiates energy into the substrate [44]. Such kind of systems have been widely studied in the last years because of their peculiar propagation features [45]. When the elastic properties of the film and the substrate are quite dissimilar, the PSAWs can eventually split, evolving into strong attenuated interfacial modes and the RW at the free surface of the layer [44,46]. The Brillouin spectra of the analyzed boron film, at different incidence angles, are shown in Fig.6. Only one acoustic mode can be measured, due to the opacity of the boron film. This mode is attributed to the PSAW mentioned above. As it can be seen the peaks are quite broad. This may be ascribed to the rough morphology of the film, in fact nano-sized irregularities act as scattering centers hindering SAW propagation. A similar behavior has already been reported in the Brillouin characterization of cluster assembled carbon films [47,48]. The analysis of the experimental spectra is conducted taking advantage of the symmetry of the Stokes/anti-Stokes parts of Brillouin spectra: the spectra presented in Fig. 6 are the geometric average of the Stokes and anti-Stokes parts. The peaks are then fitted by Gaussians (in red in Fig. 6). A clear dependence of the RW frequency on the incidence angle θ (i.e. on the exchanged wave-vector $k_\parallel$) is found. This confirms that the detected mode has a wave-vector which is parallel to the surface.

Furthermore the velocity of this mode depends on the wave-vector $k_\parallel$ because the thickness of the film is of the same order of the probed SAW wavelengths (~500 nm), in turn close to the penetration depths of the SAW. Accordingly, at lower/higher values of $k_\parallel$ the displacement field of the RW penetrates more/less deeply into the substrate and is more/less affected by the substrate properties. From the peak positions in the different spectra, the PSAW velocity can be measured, obtaining the experimental dispersion relation of the acoustic mode. The elastic constants of the film are obtained by minimizing the difference between the measured and the theoretical dispersion relations, which can be computed as a function of the following system properties: mass densities of the film and the substrate, elastic constants $C_{ij}$ of the film and the substrate, film thickness and exchanged wave-vectors. This procedure is described in detail in [49]. In the present case, due to the amorphous structure of boron films, we adopted a homogeneous continuum isotropic model for the film with the hypothesis of perfect adhesion to the substrate. For known substrate properties, and if the film mass density is known, the only independent parameters to be determined are thus $C_{11}$ and $C_{44}$. The minimization has been performed for two different values of the film mass density: 2.51 g/cm$^3$ estimable from the quartz microbalance and the bulk β-boron value of 2.35 g/cm$^3$. The two values allow to fit the experimental dispersion relation in almost indistinguishable ways, but obviously by different values of the elastic constants, as presented in Table 2. The theoretical frequencies obtained by the best fit parameters are presented in Fig. 6 by Gaussian blue peaks, whose width was taken as the instrumental one, estimated on the basis of the scattering angle and the aperture of the scattered light collection, the area of every theoretical peak is set to the value of the corresponding experimental one. Since in the PSAW the shear component prevails, our method supplies a good estimate for the shear modulus $G = C_{44}$, while leaving wider intervals in the determination of $C_{11}$.

The Young modulus E, the bulk modulus K and the Poisson's ratio ν are thus derived with larger uncertainties. The results are summarized in Table 2. A comparison with the bulk values of E, G and K of β-rhombohedral crystalline boron as well as the E values coming from nano–indentation of amorphous boron coatings deposited by cathodic arc and magnetron sputtering of B powders is also shown in the Table 2 [2, 40, 50]. The obtained results are in fair agreement with the results obtained on amorphous boron coatings. In comparison to crystalline β-rhombohedral boron the values of elastic constant are slightly lower as expected for an amorphous coating.

|  | E (GPa) | G (GPa) | K (GPa) | ν | ρ (g/cm$^3$) |
|---|---|---|---|---|---|
| Crystalline β-rhomb. B [a] [47] | 396 | 168 | 206 | 0.178 | 2.35 |
| Amorphous B coating [b] [2] | 280 | - | - | - | - |
| Amorphous B$_{0.92}$O$_{0.08}$ film [b] [40] | 273 ± 7 |  | - | - | 2.48 |
| Present study [c] | 296 ± 16 | 124 ± 8 | 159 ± 25 | 0.19 ± 0.04 | 2.51 |
| Present study [c] | 272 ± 25 | 114 ± 7 | 152 ± 32 | 0.19 ± 0.04 | 2.35 |

Table 2: Elastic moduli of different B phases measured by: [a] ultrasonic technique, [b] nano-indentation and [c] Brillouin Spectroscopy.

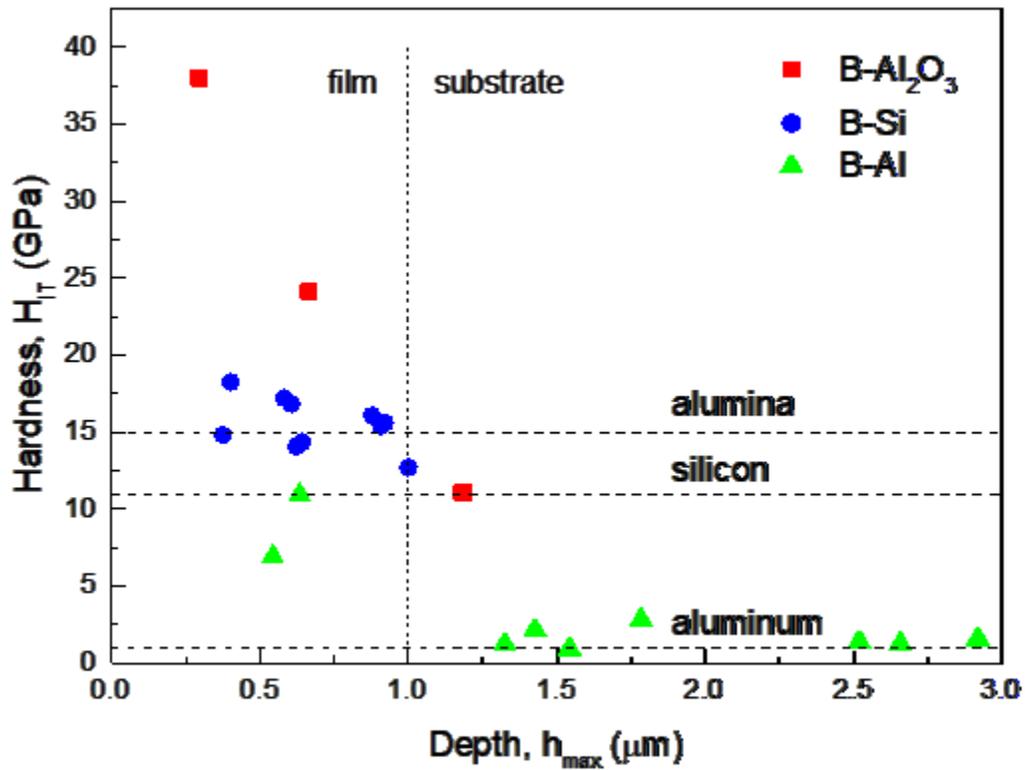

Figure 7: micro-hardness characterization of B films deposited on aluminum, silicon and alumina: hardness (GPa) versus indentation depth.

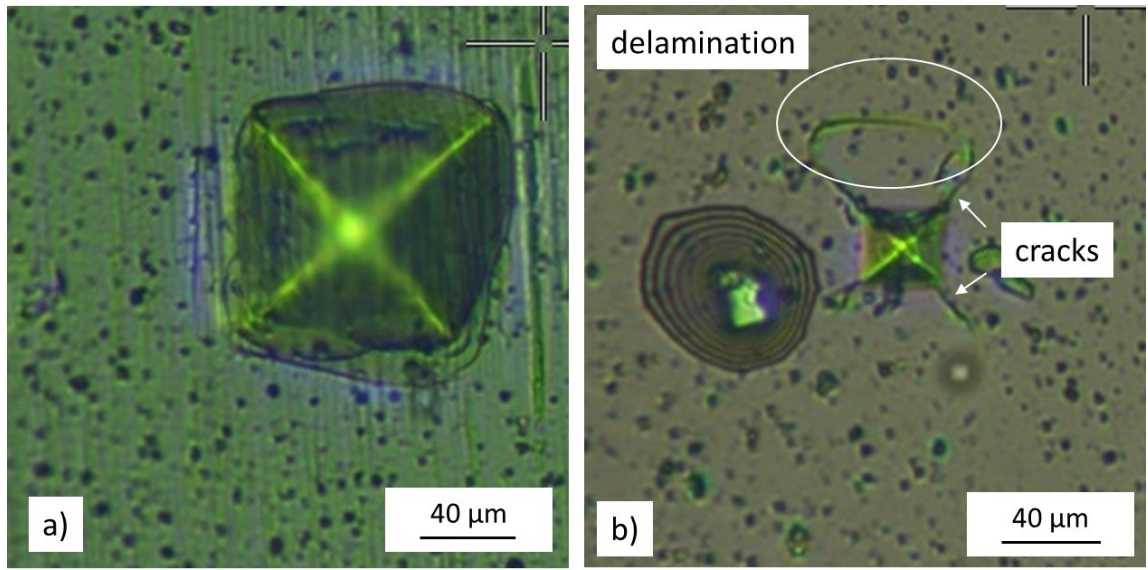

Fig. 8: Optical image of the indentation traces with 100 mN as maximum load for the B films deposited on aluminum (a) and silicon (b).

Mechanical properties of the 1 μm thick coatings have also been characterized by instrumented micro–indentation of films deposited on three different substrates, laminated aluminum, silicon and sintered alumina. The investigated substrates exhibit very different mechanical properties that allow to evaluate adhesion and cohesion of the B films during the indentation at different loads. The obtained indentation hardness ($H_{IT}$) results, see Fig.7, are affected by the behavior of the underlying substrates, especially at high penetration depths, resulting in most cases in values which are not representative of the properties of the coating alone. Nevertheless it is interesting to evaluate the behavior of the whole coating-substrate system. Due to the very low Vickers hardness of aluminum, even the smallest loads (50 mN) induce large plastic deformations. Thus the resulting B-Al system hardness is not significantly different from the typical hardness of bare aluminum (about 0.3–1 GPa). Anyway looking at the instrumented micro–indentation traces, Fig.8a, it is possible to see that the boron film follows the substrate even at high plastic deformations without detachments. Neither delamination nor particle production have been detected. Silicon substrate is much harder than aluminum (9–13 GPa [51]), flat and very brittle. Thanks to the presence of B coating the hardness of the B-Si system is higher than that of bare silicon, at the lowest penetration depth $H_{IT}$ is about 20 GPa. The coatings under load exhibit a brittle behavior, with clearly visible cracks at the imprint corners, see Fig.8b. In addition some delamination occurs. The B coating on silicon still exhibits a good cohesion, but it shows low adhesion probably due to the very small roughness of the silicon substrate. Hardness of B-alumina system versus penetration depth is visible in Fig.7. At 1 μm of penetration depth $H_{IT}$ is about 12 GPa that is compatible with the hardness of bare alumina (15–20 GPa). Lowering the penetration depth results in an increasing of the hardness. At 250 nm, about 25% of the total thickness of the sample, the measured $H_{IT}$ is 37 GPa, a value consistent with the literature data that determine Vickers hardness of boron ranging between 33 GPa and 60 GPa depending on crystalline phase and stoichiometry [1, 52]. Due to the micrometer scale roughness of the alumina substrate the resulting indentation tracks (not shown) are barely visible and it is not possible to infer any kind on information related to cohesion and adhesion of this system.

## 4. Conclusions

We deposited micron-thick amorphous boron coating, with low oxygen content, on several substrates and with surface areas up to 16 cm$^2$, exploiting the potential of nanosecond Pulsed Laser Deposition. This kind of techniques has been demonstrated to be one of the few PVD techniques able to deposit pure B coatings. Both film morphology and oxygen content are critically dependent on the energy of the impinging laser pulse. High energy regimes (>700 mJ) allow the deposition of smooth B films with low oxygen uptake (+2% respect to B target) at relatively high base pressures ($5*10^{-4}$ Pa). A detailed Raman characterization shows that, 1) the deposition base pressure that allows the growth of boron, instead of boron oxide, turns out to be $2*10^{-3}$ Pa in the present case. 2) after thermal annealing at 800 °C the crystalline order is increased showing the features of a disordered assembly of $B_{12}$ icosahedra. 3) a further annealing at 1200 °C induces the formation of a β-rhombohedral crystalline phase. The elastic behavior of amorphous boron coatings have been fully characterized by Surface Brillouin Spectroscopy. We found that the measured elastic properties are not far from that of crystalline β-rhombohedral boron. Boron coatings deposited on different substrates (Al, Si and alumina) have been also characterized by means of micro indentation, showing good adhesion properties and a hardness value of about 37 GPa. PLD performed in high energy regime has been proven to be a "simple" and effective method for the synthesis of reliable boron coatings with low oxygen content. Using this deposition strategy it is possible to grow boron coatings already at high vacuum ($10^{-3}$ Pa) reducing the pumping time and allowing the use of industrial size chambers that make possible the deposition of coatings on wide substrates. This paves the way for the use of boron films in many technology fields allowing to exploit its appealing properties. As an example the development a coating that is at the same time amorphous, hard, bio-compatible, self-lubricating and antibacterial will be of great interest in the field of applications to orthopedic implants. Another field of application of great relevance for such kind of coating could the production of pure $^{10}$B neutron detectors instead of the usual boron carbide coatings.

**Acknowledgments**

The research leading to these results has also received funding from the European Research Council Consolidator Grant ENSURE (ERC-2014-CoG No.647554).